\documentclass[fleqn,usenatbib]{mnras}

\usepackage{newtxtext,newtxmath}

\usepackage[T1]{fontenc}
\usepackage{ae,aecompl}

\usepackage{graphicx}	
\usepackage{amsmath}	
\usepackage{color}
\usepackage{bm}
\hypersetup{colorlinks,
	citecolor=[rgb]{0.1 , 0.3, 0.8}}

\newcommand{\lya}{\mbox{${\rm Ly}\alpha$}}

\newcommand{\Av}{\ensuremath{A(V)}}
\newcommand{\Avcrit}{\ensuremath{A(V)_{\rm lim}}}

\newcommand{\sigdla}{\ensuremath{\sigma_{\textsc{dla}}}}

\newcommand{\CI}{\ion{C}{i}}

\newcommand{\HI}{\ion{H}{i}}
\newcommand{\NHI}{\ensuremath{N_{\rm H\, \textsc{i}}}}

\newcommand{\logNHI}{\ensuremath{\log({\rm N_{H\, \textsc{i}}})}}
\newcommand{\logNHIcm}{\ensuremath{\log({\rm N_{H\, \textsc{i}}\: /\: cm}^{-2})}}

\newcommand{\Rcnm}{\ensuremath{R_{\textsc{cnm}}}}
\newcommand{\Rdla}{\ensuremath{R_{\textsc{dla}}}}

\title[GRB-DLA host galaxies at cosmic noon]{Modelling of Long Gamma-ray Burst Host-Galaxies at Cosmic Noon from Damped Lyman-alpha Absorption Statistics}

\author[J.-K. Krogager et al.]{
	J.-K. Krogager,$^{1, 2}$\thanks{{\tt jens-kristian.krogager@univ-lyon1.fr}}
	A. De Cia,$^{1, 3}$
	K. E. Heintz,$^{4}$
	J.~P.~U. Fynbo,$^{4}$
	L. B. Christensen,$^{4}$
	\newauthor
	G. Bj{\"o}rnsson,$^{5}$
	P. Jakobsson,$^{5}$
	S. Jeffreson,$^{6}$
	C. Ledoux,$^{7}$
	P. M{\o}ller,$^{3, 4}$
	P. Noterdaeme,$^{8}$
	\newauthor
	J. Palmerio,$^{8}$
	S. D. Vergani,$^{8,9,10}$
	and 
	D. Watson$^{4}$
	\\
$^{1}$Centre de Recherche Astrophysique de Lyon, Universit{\'e} de Lyon 1, ENS-Lyon, CNRS, UMR5574, 9 Av Charles Andr{\'e}, 69230 Saint-Genis-Laval, France\\
$^{2}$Department of Astronomy, University of Geneva, Chemin Pegasi 51, 1290 Versoix, Switzerland\\
$^{3}$European Southern Observatory, Karl-Schwarzschildstrasse 2, D-85748 Garching bei M{\"u}nchen, Germany\\
$^{4}$Cosmic Dawn Center (DAWN), Niels Bohr Institute, University of Copenhagen, Jagtvej 128, DK-2200 Copenhagen N, Denmark\\
$^{5}$Centre for Astrophysics and Cosmology, Science Institute, University of Iceland, Dunhagi 5, 107 Reykjav{\'i}k, Iceland\\
$^{6}$Center for Astrophysics, Harvard \& Smithsonian, 60 Garden Street, Cambridge, MA 02138, USA\\
$^{7}$European Southern Observatory, Alonso de C{\'o}rdova 3107, Vitacura, Casilla 19001, Santiago, Chile\\
$^{8}$Institut d'Astrophysique de Paris, Universit{\'e} Paris 6-CNRS, UMR7095, 98bis Boulevard Arago, 75014, Paris, France\\
$^{9}$GEPI, Observatoire de Paris, PSL Research University, CNRS, Univ. Paris Diderot, Sorbonne Paris Cit{\'e}, Place Jules Janssen, 92195, Meudon, France\\
$^{10}$INAF-Osservatorio Astronomico di Brera, via E. Bianchi 46, 23807, Merate, Italy
}

\begin{document}

\label{firstpage}
\pagerange{\pageref{firstpage}--\pageref{lastpage}}
\maketitle

\begin{abstract}
We study the properties of long gamma-ray burst host galaxies using a statistical modelling framework derived to model damped Lyman-$\alpha$ absorbers (DLAs) in quasar spectra at high redshift.
The distribution of \NHI\ for GRB-DLAs is $\sim$10 times higher than what is found for quasar-DLAs at similar impact parameters. We interpret this as a temporal selection effect due to the short-lived GRB progenitor probing its host at the onset of a starburst where the ISM may exhibit multiple over-dense regions. Owing to the larger \NHI, the dust extinction is larger with 29\% of GRB-DLAs exhibiting $\Av > 1$~mag in agreement with the fraction of `dark bursts'.
Despite the differences in \NHI\ distributions, we find that high-redshift $2 < z < 3$ quasar- and GRB-DLAs trace the luminosity function of star-forming host-galaxies in the same way. {We propose that} their differences may arise from the fact that {the galaxies are sampled at different times in their star formation histories, and that the absorption sight lines probe the galaxy halos differently}. Quasar-DLAs sample the full \HI\ cross-section whereas GRB-DLAs sample only regions hosting cold neutral medium.
Previous studies have found that GRBs avoid high-metallicity galaxies ($\sim$0.5\,$Z_{\odot}$). Since at these redshifts galaxies on average have lower metallicities, our sample is only weakly sensitive to such a threshold. 
Lastly, we find that the modest detection rate of cold gas (H$_2$ or \CI) in GRB spectra can be explained mainly by a low volume filling factor of cold gas clouds and to a lesser degree by destruction from the GRB explosion itself.
\end{abstract}

\begin{keywords}
\vspace{-2mm}
	gamma-ray bursts
	--- galaxies: high-redshift
	--- galaxies: statistics
\end{keywords}

\section{Introduction}

	Studying galaxy evolution via flux-limited samples tends to bias our view towards the brightest galaxies at any cosmic epoch. One way of overcoming this selection effect is to identify galaxies based on methods that probe below conventional luminosity limits. For example, it is possible to study much fainter galaxies if they are magnified by a gravitational lens in the foreground \citep[e.g.,][]{Pettini2000, Hainline2009, Christensen2012, Stark2013, Atek2018, Bouwens2022}.
	
	Alternatively, we can identify galaxies owing to their neutral gas absorption towards bright background sources \citep[damped \lya\ absorbers, DLAs, with $\logNHIcm > 20.3$;][]{Wolfe1986} or by identifying luminous transients (such as gamma-ray bursts) that hint at the existence of these faint galaxies even if their emission is not detected \citep{Tyson1988a, Natarajan1997, Wijers1998, Hogg1999, Vreeswijk2004, Jakobsson2005, Prochaska2007}.

	Gamma-ray bursts are powerful cosmic beacons that have been observed out to $z \sim 8$ \citep{Tanvir2009, Salvaterra2009}. A few host galaxies at $z\sim 6$ have been detected in emission \citep{McGuire2016}, and their afterglows can be bright enough to allow in-depth studies of the chemical abundances in their host galaxy at these high redshifts \citep{Hartoog2015b, Saccardi2023}.

	Long duration gamma-ray bursts with a duration of prompt emission longer than 2 seconds \citep[hereafter referred to simply as GRBs;][]{Kouveliotou1993} are directly associated with the deaths of massive stars \citep{Galama1998, Wijers1998, Bloom2002, Hjorth2003, Stanek2003} that trace regions of active star formation in galaxies on timescales of $\sim$10~Myr. {However, recent observations also indicate that some long-duration GRBs are associated with kilo-novae which would indicate a binary progenitor in a merger scenario \citep{Rastinejad2022}. Other recent observations also seem to hint at a more complex population of progenitors in nearby galaxies \citep{deUgartePostigo2024, Thone2024}. While some models of binary neutron star mergers predict a non-negligible fraction of mergers to occur on short timescales \citep[$<30$~Myr; e.g.,][]{Beniamini2024}, the more typical lifetimes are expected to exceed 100~Myr. Thus, at the high redshifts ($z>2$) studied in this work, the merger progenitors would contribute less to the GRB population as they require much longer time-scales.
	}
	
	Since the brightness of the GRB afterglow is not related to the luminosity of its host galaxy, we can identify and study fainter, more metal-poor galaxies that would otherwise go unseen in standard, flux-limited surveys \citep[e.g.][]{Fynbo2001b, Fynbo2008}. By obtaining spectra of the bright optical afterglow before it fades, we can furthermore study the interstellar medium (ISM) of the GRB host galaxy along the line of sight in absorption. In particular, we can characterise the column density of neutral hydrogen, \NHI, and the abundance of metals and molecules in great detail \citep{Jakobsson2006, Prochaska2007, Fynbo2009, DeCia2012, Cucchiara2015, Bolmer2019, Heintz2019c}.
	
	Despite their efficient use as cosmological probes of galaxies, GRBs may not trace the full population of star-forming galaxies in an un-biased manner; For example, an upper threshold in metallicity (or lower limit in specific star-formation rate, SFR/$M_{\star}$) has been invoked to explain the distribution of GRB host-galaxy properties \citep[e.g.,][]{Christensen2004, Vergani2015, Japelj2016, Perley2016b, Palmerio2019, Bjornsson2019, Metha2020}.
	Understanding such selection effects is vital for the interpretation of how GRBs trace galaxies as well as how they trace the cosmic star-formation rate density at the highest redshifts \citep{Kistler2009}.
	
	We can use simplified statistical models to investigate the link between the underlying galaxy population and the absorption properties that we observe. Such modelling has successfully been applied to high-redshift DLAs identified in random quasar sight lines (hereafter quasar-DLAs; \citealt{Fynbo1999, Fynbo2000, Rhodin2018, Krogager2017, Krogager2020, Krogager2020b}) and to high-redshift GRB-DLAs \citep{Fynbo2008}.
	
	If the probability of a GRB occurring in any given galaxy, and thus for that galaxy to be identified as a GRB host, is proportional to the star formation rate (SFR), as posited by \citet{Porciani2001}, then the ensemble of GRB host galaxies should be sampled uniformly from the overall population of star-forming galaxies weighted by their SFR \citep{Fynbo2008}.
	
	Quasar-DLAs on the other hand sample the star-forming galaxy population weighted by their cross-section of DLA gas, $\sigdla$, where $\NHI > 2\cdot 10^{20}$~cm$^{-2}$ \citep{Wolfe1986}. We assume that this projected DLA cross-section is proportional to luminosity \citep{Krogager2020}. At these high redshifts ($z\sim 2$), the passive galaxy population contributes no more than 20\% of the total stellar mass \citep[e.g.,][]{Santini2022}, and since the high-column density \HI\ cross-section is expected to be suppressed for passive galaxies, we therefore neglect the contribution of passive galaxies in this model.
	Assuming that SFR scales directly with the UV luminosity as well, \citep[][for unobscured star formation; see discussion in Sect.~\ref{discussion:dark}]{Kennicutt1998}, GRB-DLAs and quasar-DLAs both sample the galaxy population weighted by luminosity. We would therefore naively expect the host galaxies of high-redshift GRB-DLAs and quasar-DLAs to be drawn from the same underlying population of star-forming galaxies, i.e., they sample the luminosity function in the same way.
	
	The aim of this paper is to test this simple scenario by comparing the observed properties of GRB-DLAs to the model by \citet{Krogager2020} and \citet{Krogager2020b}. Using the two-phase model of the neutral medium implemented by \citet{Krogager2020b}, we assume that GRBs only arise in parts of galaxies where the cold and dense neutral medium can be maintained in pressure balance, since the cold dense gas is needed for the formation of stars. Given the way quasar sight lines randomly probe the galactic environment, quasar-DLAs on the other hand probe predominantly the more extended warm neutral medium.
	We then compare the model predictions to the observed distributions of \NHI, metallicity, \Av, and impact parameters of GRB-DLAs. In this work, we only consider high-redshift bursts ($z \gtrsim 2$) for which \NHI\ can be measured directly from ground-based spectroscopy via the Lyman-$\alpha$ transition.

	The paper is organized as follows: We first describe the compilation of data from the literature in Sect.~\ref{data}. In Sect.~\ref{model}, we present an overview of the statistical model used in our work and how this is applied to model GRB host galaxies before presenting our results of the model comparison in Sect.~\ref{results}. In Sect.~\ref{discussion}, we offer a discussion of how GRBs trace star-forming galaxies in light of our modelling results. Lastly, in Sect.~\ref{summary}, we provide a short summary of our findings.
	
	Throughout this paper, we use the following cosmological parameters: a flat $\Lambda$CDM cosmology with $H_0=68\, \mathrm{km s}^{-1}\mathrm{Mpc}^{-1}$ and $\Omega_{\textsc{m}} = 0.31$ \citep{Planck2016}.

\section{Sample Selection}
\label{data}

	Since the model to which we compare our data {has been tuned to absorption properties at redshift $z = 2.5$}, we only consider bursts in the redshift range $2 \lesssim z \lesssim 3.5$.

	\subsection{Metallicity and dust extinction}
	\label{data:metals}
	We collect a sample of absorption metallicities and dust extinction measurements from GRB afterglow spectroscopy with intermediate (or higher) spectral resolution. We include measurements from VLT/X-shooter, VLT/UVES, Keck/ESI and Keck/HIRES in order to have robust metallicity measurements.

	{
	In order to constrain the gas-phase metallicity the spectroscopic follow-up of the afterglow has to be of sufficient quality to constrain the column densities of the narrow metal lines. This therefore introduces a bias against dust-obscured bursts \citep[see also][]{Perley2016b}. This is directly evident in the distribution of \Av\ in our sample compared to the more complete study by \citet{Covino2013} who find values of \Av\ up to $\sim3$~mag. This incompleteness is taken into account in our modelling, see Sect.~\ref{sect:model-corrections}.
	}
	
	The compilation of metallicities and dust extinction for the total of 24 GRBs is given in Table~\ref{tab:metals}. This sample is referred to as the `\emph{GRB metal sample}' in the remainder of this paper.

	\subsection{Neutral hydrogen column density}
	
	{In order to improve the statistics on \NHI\ measurements and to investigate possible biases related to the metallicity determinations, we collect measurements of \NHI\ from the comprehensive work by \citet{Tanvir2019}.}
	Since we only consider GRB-DLAs in this work, we restrict the sample by \citet{Tanvir2019} to those absorbers with $\logNHIcm > 20.3$.
	{When comparing the Tanvir et al. sample to our GRB metal sample there appears to be a slight lack of high-\NHI\ absorbers in the metal sample, which could indicate a dust-obscuration bias as mentioned above.
	However, the difference is not statistically significant, as evidenced by the $p$-value of the two-sample Kolmogorov–Smirnov (KS) test $p = 0.72$.
	}

	\subsection{Impact Parameters}

	The sample of impact parameters, i.e., the projected distance from the GRB explosion site to the luminosity-weighted centre of the galaxy, used in this paper has been taken from the work by \citet{Lyman2017}.
	{
	These authors associate host galaxies to GRBs using near-infrared imaging data from the \emph{Hubble Space Telescope} (i.e., rest-frame optical).} We consider bursts down to a slightly lower redshift limit ($1.5 < z < 3.5$) than for the metallicities, since the sample is otherwise too small for a meaningful comparison. In total there are 9 bursts that meet our criteria.
	All impact parameters have been corrected for differences in assumed cosmology.

	\citet{Lyman2017} report non-detections of impact parameters for 6 out of 15 bursts that meet our redshift criteria. Hence, $40 \pm 13$\% of GRB host galaxies at $z>1.5$ are not detected down to a luminosity limit of $M \gtrsim -18$ (	{rest-frame $V$- or $R$-band depending on the redshift}; see their fig.~5 and sect.~4.2). This may introduce a bias in the distribution of impact parameters that will be discussed in more detail in Sect.~\ref{results}.
	
	{
	We also compare to the work by \citet[][see also \citealt{Bloom2002, Fruchter2006}]{Blanchard2016} who study the associated host galaxies in a mix of near-infrared and optical photometric bands (rest-frame UV to optical). We again only consider bursts in the redshift range $1.5 < z < 3.5$, which gives a total sample of 32 bursts. Out of these, 6 bursts have no associated host. The fraction of non-detected bursts ($19\pm6$\%) is thus lower than what \citet{Lyman2017} infer but consistent within the rather large uncertainties due to the low number of bursts.
	
	In some cases the host galaxies associated by Blanchard et al. seem to be biased towards brighter and larger galaxies in the field. One example of this is seen for the burst 080319C at $z=1.95$ which is studied by both teams. Blanchard et al. associate a bright spiral galaxy at a projected distance of 7~kpc as the host, whereas Lyman et al. assign a much fainter source as the host at a projected distance of 0.3~kpc. In this work, we compare our model to the measurements of both studies in order not to give preference to any particular sample.
	}

\begin{table}
\centering
\renewcommand{\arraystretch}{1.10}	
\caption{Compilation of GRB {\it metal sample} from afterglow absorption spectroscopy of bursts in the redshift range $2 < z < 3.5$.}
\label{tab:metals}
\begin{tabular}{lccccr}
\hline
GRB & $z_{\rm abs}$ & [X\,/\,H] & X & \Av & Ref.  \\\hline
000926  & 2.0380 & $-$0.11 & Zn & 0.38 & (1, 9)    \\
030226  & 1.9870 & $-$1.05 & Fe & 0.00 & (2)     \\
050820A & 2.6150 & $-$0.39 & Zn & 0.27 & (4, 2, 6, 9)  \\
050922C & 2.1990 & $-$2.09 & Si & 0.00 & (4, 3)      \\
071031  & 2.6920 & $-$1.76 & Zn & 0.00 & (4)         \\
080413A & 2.4330 & $-$1.63 & Zn & 0.00 & (4)         \\
081008  & 1.9685 & $-$0.52 & Zn & 0.08 & (5, 6) \\
090809A & 2.7373 & $-$0.86 & Zn & 0.11 & (7)  \\
090926A & 2.1069 & $-$1.97 & Zn & 0.03 & (7)  \\
111107A & 2.8930 & $-$0.74 & Si & 0.15 & (7)  \\
120327A & 2.8143 & $-$1.49 & Zn & 0.05 & (7)  \\
120716A & 2.4874 & $-$0.71 & Zn & 0.30 & (7)  \\
120815A & 2.3582 & $-$1.45 & Zn & 0.19 & (7)  \\
121024A & 2.3005 & $-$0.76 & Zn & 0.26 & (7)  \\
130408A & 3.7579 & $-$1.48 & Zn & 0.12 & (7)  \\
141028A & 2.3333 & $-$1.64 & Si & 0.13 & (7)  \\
141109A & 2.9940 & $-$1.63 & Zn & 0.16 & (7)  \\
150403A & 2.0571 & $-$1.04 & Zn & 0.12 & (7)  \\
151021A & 2.3297 & $-$0.98 & Zn & 0.20 & (7)  \\
160203A & 3.5187 & $-$1.31 & S  & 0.10 & (7)  \\
161023A & 2.7100 & $-$1.23 & S  & 0.09 & (7)  \\
170202A & 3.6456 & $-$1.28 & S  & 0.08 & (7)  \\
181020A & 2.9379 & $-$1.50 & Zn & 0.27 & (8)  \\
190114A & 3.3764 & $-$1.16 & Zn & 0.36 & (8)  \\
\hline
\end{tabular}

{\bf References:} (1) \citet{Savaglio2003}; (2)~\citet{Prochaska2007}; (3)~\citet{Piranomonte2008}; (4)~\citet{Ledoux2009}; (5)~\citet{DElia2011}; (6)~\citet{Wiseman2017}; (7)~\citet{Bolmer2019}; (8)~\citet{Heintz2019c}; (9)~\citet{Zafar2019}

\end{table}

\section{Statistical Modelling}
\label{model}
	
	The model used in this work is based on the modelling approach by \citet{Fynbo2008}, which has been developed further to include a statistical prescription for \NHI\ \citep{Krogager2020} as well as a two-phase description of the neutral gas as either a warm or cold phase \citep{Krogager2020b}. We here offer a short summary of the model framework. For details regarding the model implementation, see \citet{Krogager2020} and \citet{Krogager2020b}.
	
	The model has been designed to study the properties of quasar-DLAs in a simplified manner. These absorbers are selected based on their projected \HI\ cross-section with the probability density $P_{\textsc{qso-dla}}(L) \propto \sigdla(L) \ \phi(L)$, where $\sigdla(L)$ is the effective cross-section of DLAs and $\phi(L)$ is the UV luminosity function.

	\citet{Krogager2020} use a simple circular projected cross-section, such that $\sigdla(L) = \pi R_{\textsc{dla}}(L)^2$. The typical radial extent of \sigdla\ is further assumed to scale with luminosity motivated by the tight mass--size relation for \HI\ \citep{Stevens2019}. We use a Holmberg relation of the form: $R_{\textsc{dla}}(L) = R^*_{\textsc{dla}} (L/L^*)^t$ where $R^*_{\textsc{dla}}$ is the radial extent of the DLA cross-section for an $L^*$ galaxy. \citet{Krogager2020} obtain a value of $t=0.5$, i.e., $\sigdla(L) \propto L$.
	The absolute scaling of $R_{\textsc{dla}}(L)$, which in turn determines the typical impact parameters of quasar DLAs, is obtained by normalizing the cross-section to the DLA incidence rate, ${\rm d}n_{\textsc{dla}}/{\rm d}z$ \citep{Zafar2013b}.
	Galaxies are then sampled from the luminosity function weighted by \sigdla, and an impact parameter, $b$, for each randomly sampled galaxy is drawn with a probability $P(b) \propto b$ for $b\leq R_{\textsc{dla}}(L)$.

	Each galaxy in the model population is assigned a global metallicity, $Z_0$, following a metallicity--luminosity relation: $\log(Z_0) = \log Z_0^* + \beta \times M_{\textsc{uv}}$ with $\beta = 0.2$.
	A radial metallicity gradient is assumed to obtain the absorption metallicity, $Z_{\rm abs}$, at the radial position given by $b$. This gradient is taken to be luminosity dependent with a variable power-law index:
	$\gamma = \gamma^*\, L^{-0.5}$ following \citet{Boissier2001} with $\gamma^* = -0.019$~dex~kpc$^{-1}$ \citep{Krogager2020}.
	However, at the small physical scales probed by the GRBs in this model, the metallicity gradient is negligible.
	
\begin{figure}
	\includegraphics[width=0.48\textwidth]{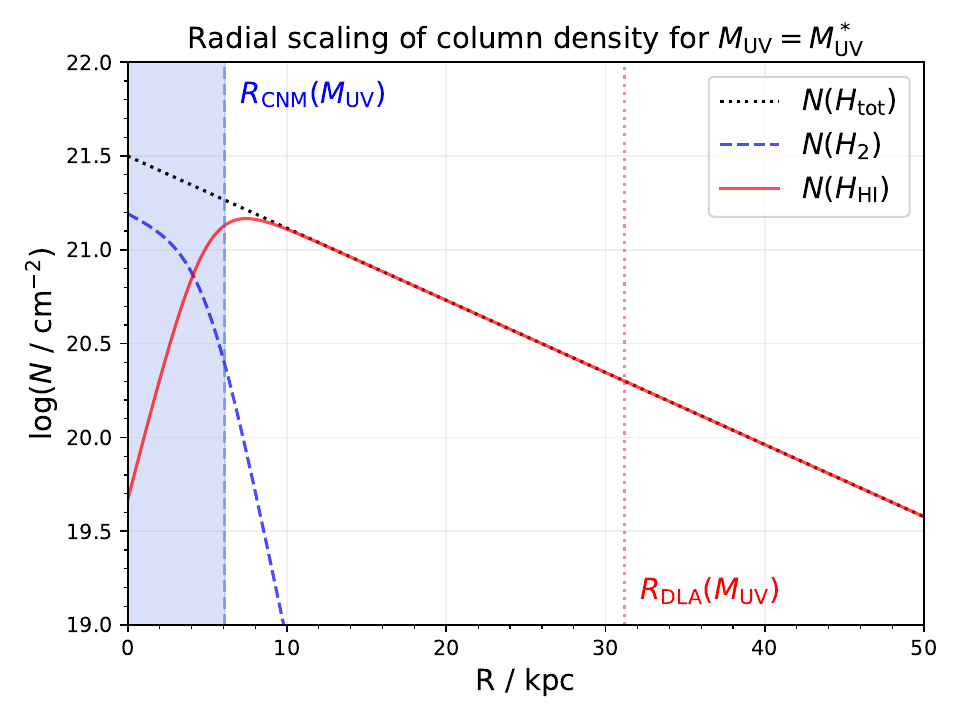}
	\caption{Average column density of total H, H$_2$ and \HI\ as a function of galactic radius for an $L^*$ galaxy in our model.  The radius of \Rdla\ (red dotted vertical line) scales with host galaxy luminosity as shown in Fig.~\ref{fig:cross-section} and denotes the outer extent of the randomly drawn impact parameters in our model. 
	Within \Rcnm\ (blue dashed vertical line) defined by $P > P_{\rm min}(Z)$ \citep{Wolfire2003} the dense CNM clouds are stable in pressure equilibrium with the WNM giving rise to H$_2$ and \CI\ absorption. The \Rcnm\ from our model coincides with the transition from the \HI-dominated phase to the H$_2$-dominated phase. GRBs are expected to arise within \Rcnm\ (light blue shaded region) as star formation only occurs in these cold dense clouds. The central column density of total H is assumed to be constant for all galaxies motivated by local observations \citep{Bigiel2012}.
	}
	\label{fig:radial_logN}
\end{figure}

	Together with the absorption metallicity, the model further assigns a total hydrogen column density, which is subsequently split into \HI\ and H$_2$ column densities \citep[following][see Fig.~\ref{fig:radial_logN}]{Blitz2006,Bigiel2012}. \NHI\ is then used to calculate the rest-frame optical dust-extinction along the line of sight, \Av, assuming a constant dust-to-metal ratio \citep{Krogager2019, Zafar2019}: $\log(\Av) = \log(Z_{\rm abs}) + \logNHIcm + \kappa$, with $\kappa = -21.4$.

\begin{figure}
	\includegraphics[width=0.48\textwidth]{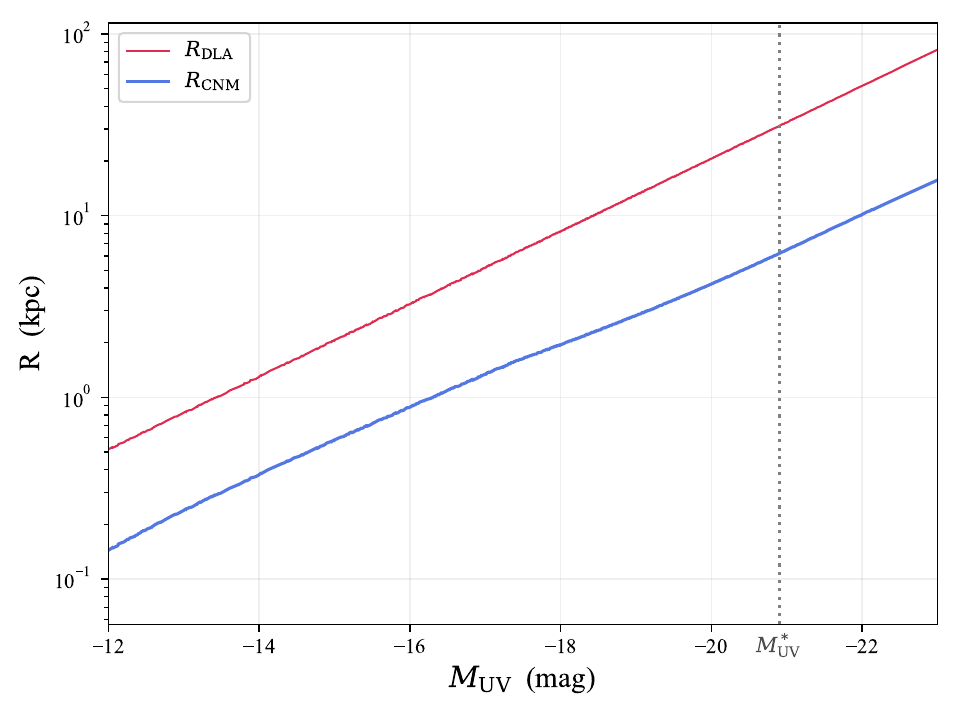}
	\caption{
	Radial extent of the projected DLA and CNM cross-section as a function of host galaxy luminosity in our model.
	}
	\label{fig:cross-section}
\end{figure}

	\citet{Krogager2020b} include a pressure-based two-phase description of the neutral gas following the canonical model by \citet{Field1969} of a warm and cold neutral medium (WNM and CNM, respectively). The two phases are modelled by a metallicity-dependent pressure threshold, $P_{\rm min}(Z)$, above which the neutral gas can exist in a stable cold phase \citep{Wolfire1995, Wolfire2003}. A radial pressure gradient \citep[see][]{Elmegreen1994} as a function of luminosity is included in the model, giving rise to a characteristic radius, $R_{\textsc{cnm}}$, within which the pressure is large enough to sustain a stable CNM, i.e., $P(r) > P_{\rm min}$. The radial extent of these two phases, $R_{\textsc{dla}}(L)$ and $R_{\textsc{cnm}}(L)$, are shown in Fig.~\ref{fig:cross-section} as a function of host galaxy luminosity. Since stars only form out of the dense and cold gas phase, we expect star formation to occur predominantly on spatial scales within $R_{\textsc{cnm}}$. We therefore restrict our GRB model to only consider sight lines that probe this CNM region (Fig.~\ref{fig:radial_logN}). Quasar-DLAs, on the other hand, probe the full extent of neutral gas within $R_{\textsc{dla}}$ and less frequently within $R_{\textsc{cnm}}$ due to the smaller projected cross-section. The relative cross-sections of DLA and CNM gas defined this way reproduces well the observed scarcity of H$_2$ detections in DLAs \citep[e.g.,][]{Ledoux2003, Jorgenson2010, Balashev2018, Krogager2020b}.

	We highlight that our statistical model only considers the on-sky projected gas column density distribution. The column density and metallicity of a given absorption sightline is drawn from a one-dimensional radial distribution function, such as shown in Fig.~\ref{fig:radial_logN}, based on the randomly assigned impact parameter.

	\subsection{The baseline GRB model}
	
	As a starting point for our model comparison, we make the simplifying assumption that GRBs directly trace star formation rate \citep[e.g.,][]{Porciani2001, Robertson2012}, which is assumed to scale with the UV luminosity \citep{Kennicutt1998}; Hence, $P_{\textsc{grb-dla}}(L) \propto {\rm SFR}\ \phi(L) \propto L\ \phi(L) \propto P_{\textsc{qso-dla}}(L)$. In other words, GRB-DLAs and quasar-DLAs in our model trace the luminosity function of high-redshift star-forming galaxies in a similar way. In our baseline model, their differences therefore arise from the fact that quasar-DLA sight lines are selected randomly within the full extent of neutral gas, i.e., within $R_{\textsc{dla}}(L)$, whereas GRB-DLAs only trace the regions of recent star formation, i.e., within $R_{\textsc{cnm}}$. This is similar to the cartoon illustration by \citet[][see their fig. 1]{Prochaska2007}.
	
	Since the GRB explosion arises within the host galaxy ISM, the absorption line of sight does not probe the full volume of the neutral gas in the same way as quasar sight lines do. We therefore calculate a statistical geometric correction to the GRB sight lines to take into account this diminished column density in front of a GRB as compared to a quasar that hits within \Rcnm.
	For uniformly distributed burst locations, the average value of this geometric correction is exactly 0.5, i.e., GRB-DLA sight lines should have half the total hydrogen column density compared to quasar-DLAs.

	A full 3-dimensional modelling of the complex density field is beyond the scope of this simplified model. We therefore scale down the column densities of GRB sight lines by a factor of $q$ drawn randomly within the interval $[0.1; 0.9]$ with $<q>\, =\, 0.5$. We still only consider sight lines in the model with $\logNHIcm > 20.3$ after the geometric correction, given our sample selection criterion. The statistical model explained above with the additional geometric correction will be referred to as our baseline GRB host galaxy model – or `baseline GRB model'.

	\subsection{GRB model corrections}
	\label{sect:model-corrections}

	A direct comparison of our GRB data to the baseline GRB model described above yields acceptable agreement for the metallicity and impact parameters, when assuming that GRBs only probe the CNM regions and including the geometric correction. However, the distributions of \NHI\ and \Av\ are both significantly under-estimated for the same model.
	For reference, these baseline results are shown as blue lines in Figs.~\ref{fig:CDF} and \ref{fig:impact-pars}.
	As the baseline model already takes into account the smaller average impact parameters of GRB-DLAs, the observed excess of \NHI\ is therefore not caused by higher densities at smaller galactic radii (see also Sect.~\ref{discussion:dlas}). Instead, we must introduce one extra degree of freedom in our model to reproduce all observations simultaneously. This additional parameter represents an overall scaling of the average \NHI\ along GRB absorption sight lines. The interpretation of this additional parameter will be discussed in Sect.~\ref{discussion:overdensity}.

	The density scaling of \HI\ is included via the parameter $\Delta \logNHI$ which is the average over-density of \NHI\ within the CNM region of the host galaxy, i.e., a constant offset to \logNHI\ for GRB sight lines. We note that the over-density of \HI\ might depend on other gas properties such as the metallicity. However, lacking any physically motivated scaling between the density and metallicity, we have ignored such higher-order complexities in our analysis. However, we already note that a constant offset to \NHI\ rather than \logNHI\ would not provide a solution to the problem, as it would introduce a sharp edge to the distribution of \logNHI\ around this value.

	With an additional contribution to \NHI, the amount of dust extinction is also expected to increase for a fixed dust-to-gas ratio. We therefore similarly increase \Av\ for GRB sight lines by scaling the initial \Av\ to the new value of \NHI\ for any given GRB sightline. Lastly, we change the implementation of the dust bias, which was modelled following \citet{Krogager2019} for quasar-DLAs. Since GRB detection is not susceptible to the complex colour {and magnitude} selection criteria that quasar-DLAs are, we simply include a binary selection criterion; GRBs with $\Av\ > \Avcrit$ are too obscured to yield metallicity measurements in absorption and will therefore be excluded in our model distributions \citep[see][]{Fynbo2009}. The value of $\Avcrit$ is freely variable in order to best reproduce the observed distribution of $\Av$ for GRBs as mentioned in Sect.~\ref{data:metals}. 
	
	The two parameters, \Avcrit\ and $\Delta \logNHI$, are fitted to the observed distributions for the GRB metal sample. The best-fit parameters are $\Delta \logNHI = 1.06\pm0.09$ (i.e., the average \HI\ density is 9--14 times higher) and $\Avcrit = 0.44\pm0.05$~mag, see Fig.~\ref{fig:NHI_dust_fit}. This best-fit model adaptation of the baseline model is referred to as the `GRB model' in what follows.

	To test the proposed metallicity thresholds for GRB progenitors we have included calculations for two such thresholds: $Z < 0.7\,Z_{\odot}$ \citep{Palmerio2019} and $Z < 0.35\,Z_{\odot}$ \citep{Metha2020}.

\subsection{Model Results}
\label{results}

\begin{figure}
	\includegraphics[width=0.48\textwidth]{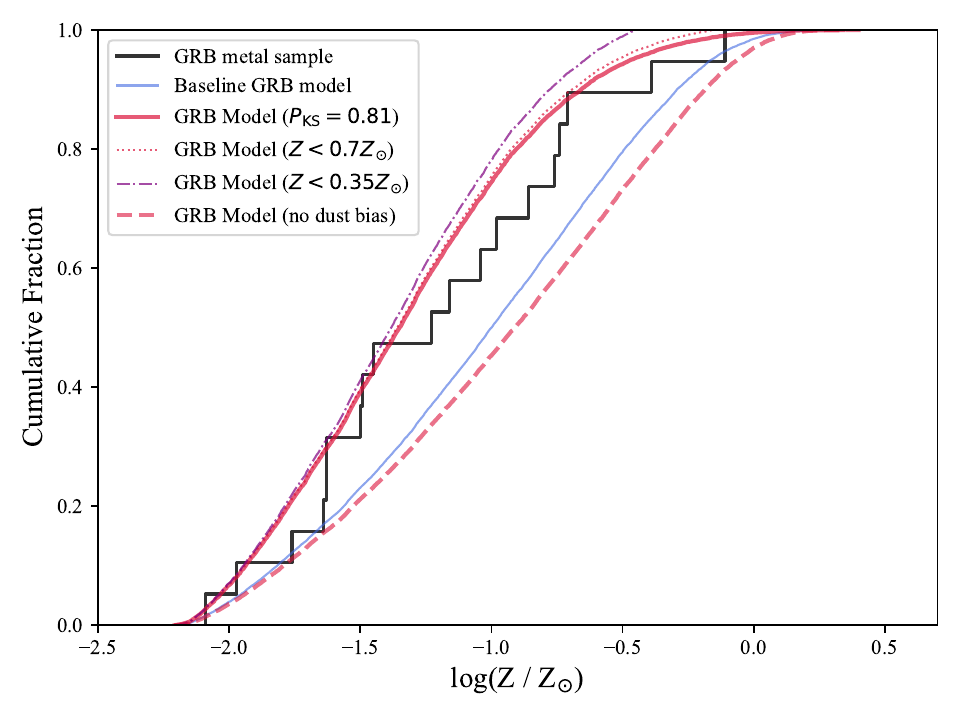}
	\includegraphics[width=0.48\textwidth]{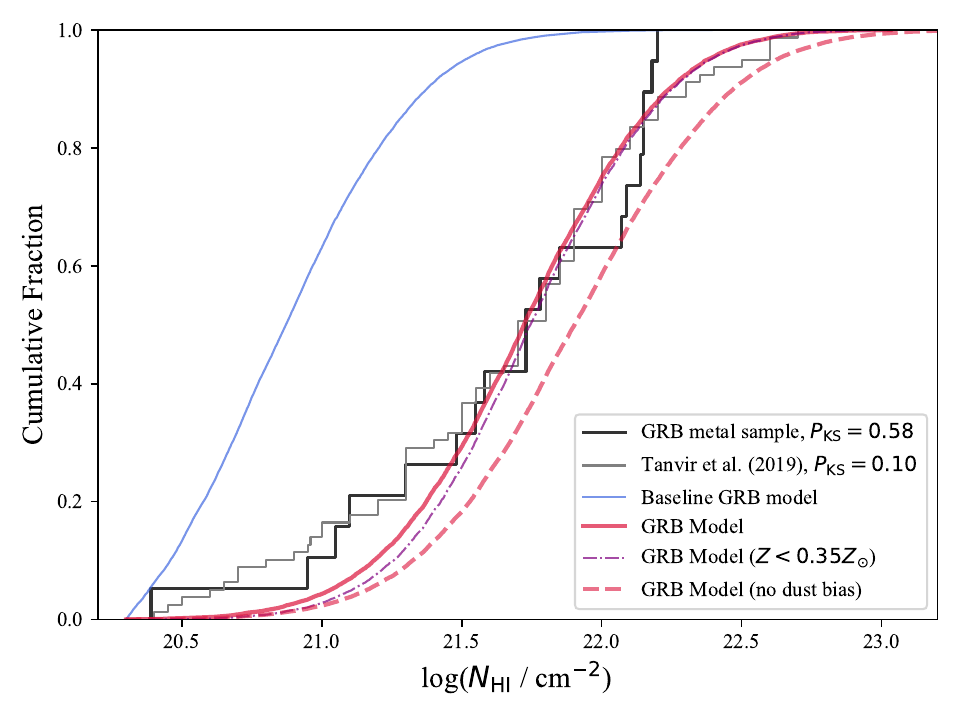}
	\includegraphics[width=0.48\textwidth]{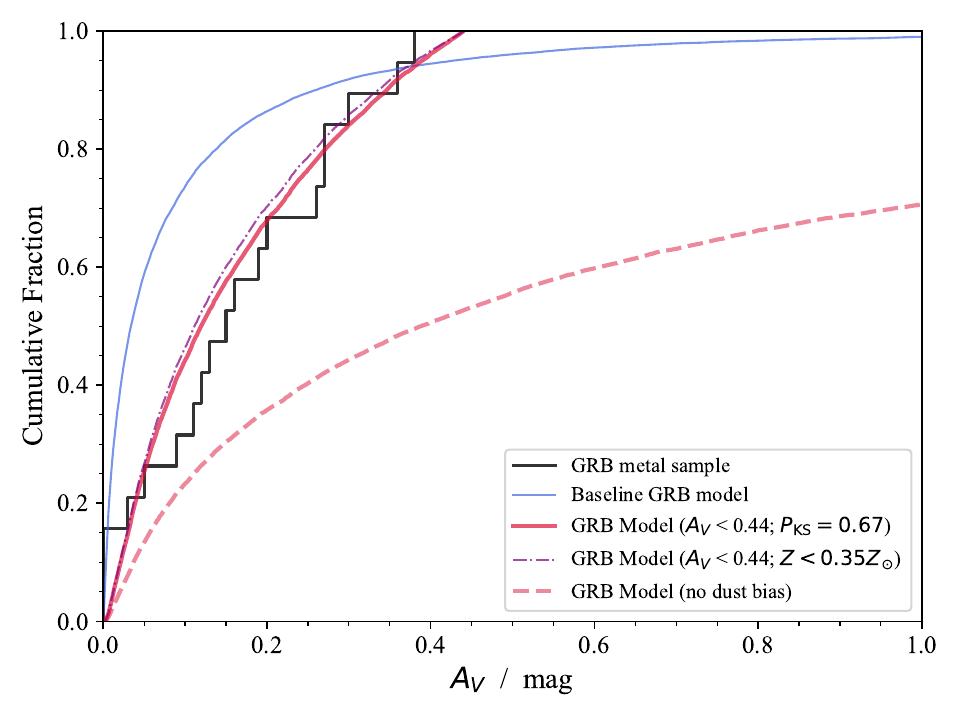}
	\caption{Cumulative distribution functions (top to bottom) of absorption metallicity from afterglow spectroscopy, \HI\ column density from afterglow spectroscopy, and rest-frame optical dust extinction, \Av. The red line shows the best model allowing for an increase in average density of the CNM and including GRB-specific dust bias. The red, dashed line shows the same model but without the dust bias. The blue line in all panels indicates the baseline model before the density and dust-bias tuning. The KS test $p$-values are calculated for the GRB model (red lines) compared to the metal sample. In the middle panel, we also give the KS $p$-value for the Tanvir et al. sample for comparison though the fit was performed using the GRB metal sample. In each panel, we further show a GRB model with an imposed 0.35$Z_{\odot}$ metallicity threshold (purple dash-dotted line). The higher-metallicity threshold of 0.7$Z_{\odot}$ is not included in the middle and bottom panels, as it is indistinguishable from the GRB model.}
	\label{fig:CDF}
\end{figure}

	Fig.~\ref{fig:CDF} shows the observed distributions of metallicity, \NHI\ and \Av\ for the GRB metal sample in black. The additional GRB sample by \citep{Tanvir2019} used for comparison is shown in grey. The model predictions for $\sim$15,000 CNM sight lines drawn randomly from the parent galaxy population are shown as the thick, red lines and the $p$-values based on a KS-test comparing the model to the observed samples is indicated in the legend of each panel. For comparison, we also show the baseline model distribution for GRB-DLAs (without the additional density parameter) as thin, blue lines. The intrinsic model distribution without taking an observational dust-bias into account is shown by the red, dashed lines. In the top panel of Fig.~\ref{fig:CDF}, we also show the distributions assuming the two proposed metallicity thresholds of 0.7 \citep{Palmerio2019} and 0.35 \citep{Metha2020} Solar as the red, dotted and purple, dash-dotted lines, respectively. Note that the red, dotted lines are not included in the other panels as they are indistinguishable from the thick red line.

\begin{figure}
	\centering
	\includegraphics[width=0.48\textwidth]{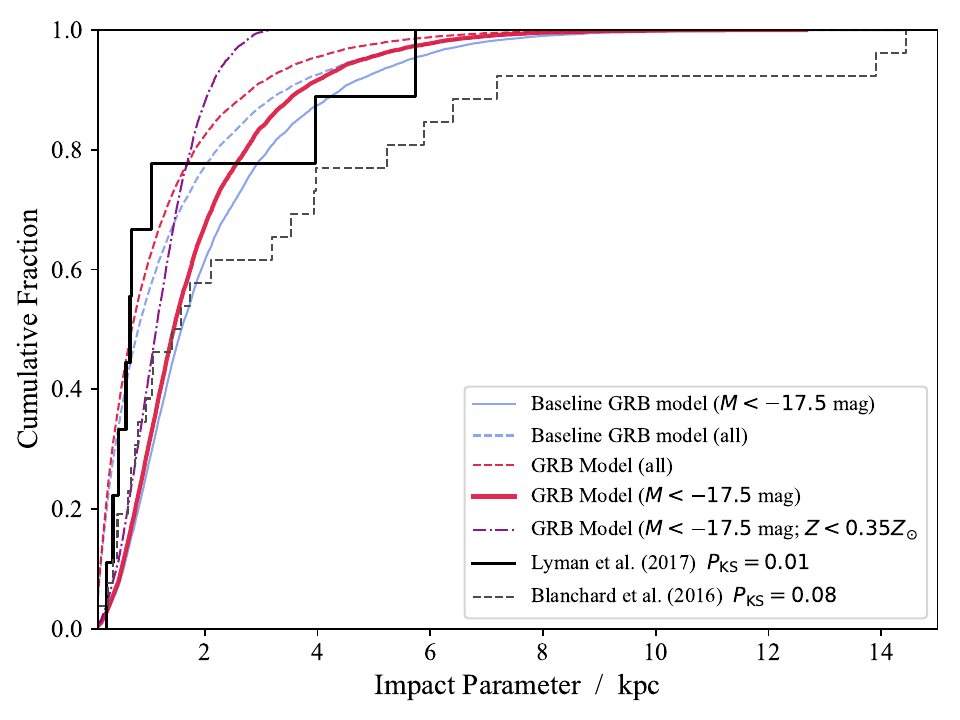}
	\caption{Cumulative distribution functions of observed impact parameters of GRBs from \citet{Lyman2017} in black and from \citet{Blanchard2016} as the grey, dashed line. The red lines show the GRB model distribution, and the thin, blue lines show the baseline model results. Solid lines refer to the magnitude limited sample (see text) and the dotted lines indicate the distributions when considering all host galaxies. The purple dash-dotted line includes a metallicity threshold.}
	\label{fig:impact-pars}
\end{figure}

\begin{figure}
	\centering
	\includegraphics[width=0.45\textwidth]{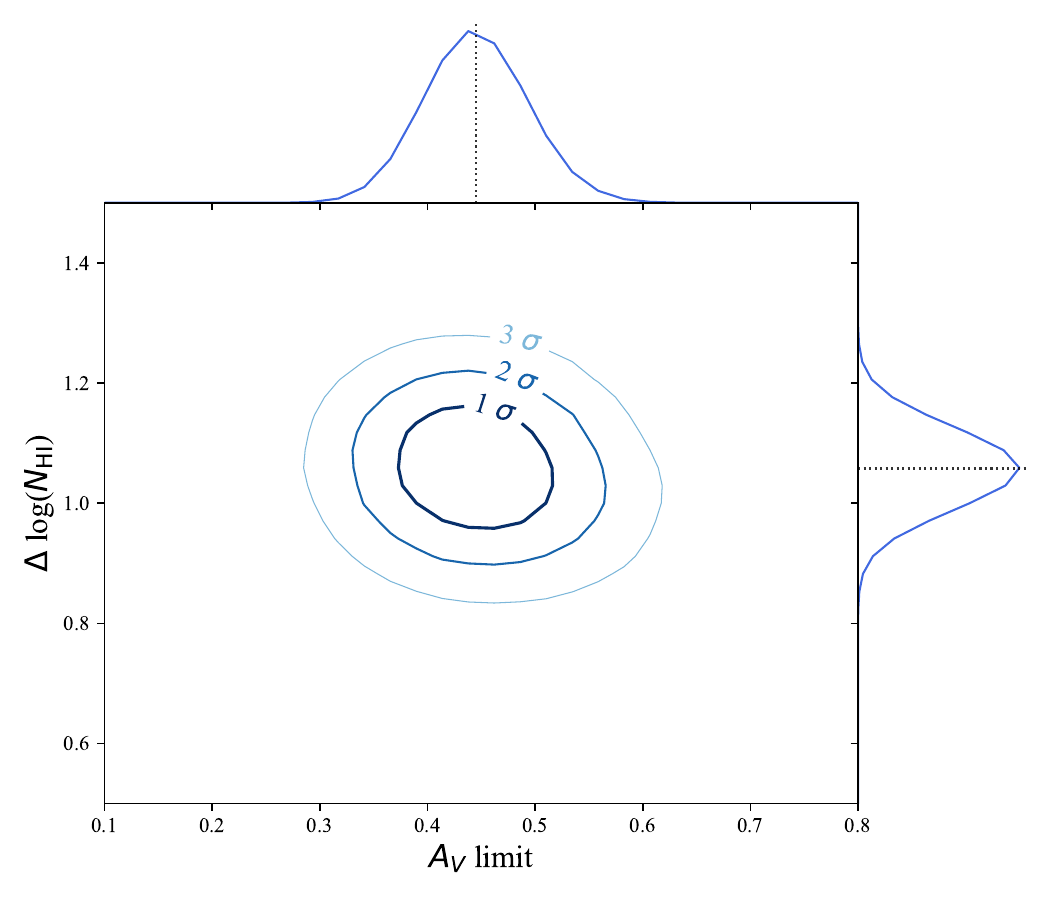}
	\caption{Constraints on $\Delta \logNHI$ and \Avcrit\ parameters. $\Delta \logNHI$ parametrizes the over-density of \HI\ along GRB sight lines compared to quasar sight lines through the same regions of galaxies.
	\Avcrit\ quantifies the effects of a dust obscuration bias since we only consider GRB afterglows with absorption-derived metallicities.}
	\label{fig:NHI_dust_fit}
\end{figure}

\begin{figure}
	\includegraphics[width=0.48\textwidth]{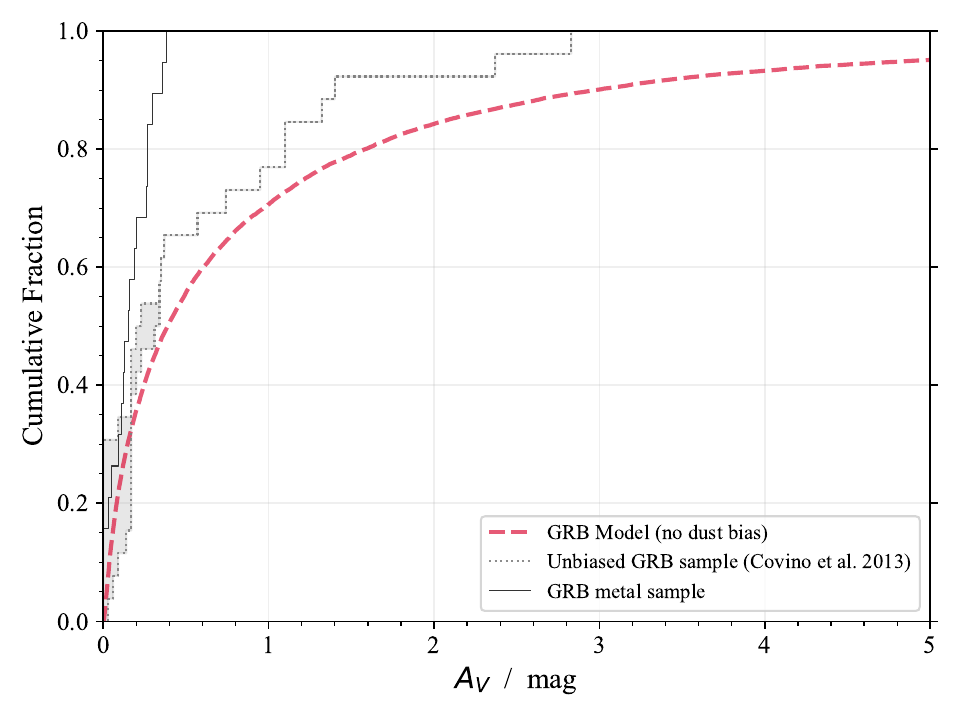}
	\caption{Comparison of unbiased \Av\ for high-redshift GRBs by \citet{Covino2013}.  The gray shaded region indicates the range allowed by the upper-limits given by these authors. The GRB model is shown as the dashed red line (same as bottom panel of Fig.~\ref{fig:CDF}) without the \Avcrit\ criterion in order to reproduce an intrinsic sample of GRB-DLAs without the additional requirement that the afterglow be bright enough for spectroscopic analysis. For comparison, we also show the GRB metal sample as the thin solid line.
	}
	\label{fig:unbiased_Av}
\end{figure}

	We observe a good agreement between the absorption properties from the GRB metal sample when compared to the full GRB model for which we include one additional free parameter (the additional \HI\ density). Moreover, the higher density of neutral gas needed to reproduce the observed \NHI\ distribution leads to a high fraction of highly obscured bursts which would not enter the metal sample due to the need for good spectroscopic follow-up. This is illustrated in Fig.~\ref{fig:unbiased_Av} where we compare the sample of \Av\ measurements by \citet{Covino2013} to the intrinsic distribution from the GRB model. The sample by Covino et al. is not restricted by the need to obtain detailed absorption measurements, and is thus more representative and less biased than the GRB metal sample. This is reflected in the much larger \Av\ measurements obtained in their sample. We find good agreement between our model without the \Av\ limit and the unbiased sample by \citet{Covino2013} at moderate levels of extinction (for $\Av \lesssim 1$~mag). Our model predicts a fraction of 29\% of so-called `dark bursts' when adopting the definition by these authors, i.e., $\Av > 1$~mag.

	In Fig.~\ref{fig:impact-pars}, we show the observed distribution of impact parameters from both samples described in Sect.~\ref{data} compared to the baseline model and the GRB model. In order to carry out a fair comparison, we restrict our models to similar host galaxy luminosities as the observational magnitude limit\footnote{It is here assumed that all GRBs would intrinsically be bright enough to be detected if there were no dust along the line of sight.}. This luminosity comparison of the host galaxy is not straightforward due to different wavelength ranges probed; The sample by \citet{Lyman2017} is restricted to the rest-frame optical ({\it HST/F160W}, $\sim$4400~\AA\ at $z=2.5$) and the sample by \citet{Blanchard2016} is a mixture of rest-frame optical and UV, whereas our model only produces rest-frame luminosities at 1700~\AA. To simplify the calculation, we assume a flat spectral shape (in $F_{\nu}$) between the UV and the optical. Yet, even a small amount of dust optical extinction can severely suppress the UV flux due to the steep extinction curves. We therefore need to correct the UV luminosities for dust-obscuration. Lacking a global dust attenuation model, we use the model \Av\ along the line of sight as a proxy of the overall attenuation of the host galaxy, though there is significant scatter between these two quantities \citep[e.g.,][]{Kruhler2011}. The rest-frame \Av\ is converted to extinction at 1700~\AA\ by multiplying by a factor of 4 assuming an average SMC extinction curve \citep{Gordon2003}. We then apply a luminosity cut to these dust-corrected model luminosities using a limit of $-17.5$~mag as inferred roughly from \citet[][their fig. 5]{Lyman2017}. Furthermore, we include a cut on $\Av < 3$~mag motivated by the maximum value observed by \citet{Covino2013} to take into account the bursts that have no optical/near-infrared afterglow identified and therefore cannot have an accurate impact parameter measurement.
	
	We show this luminosity restricted model in Fig.~\ref{fig:impact-pars} as the solid, red line. The same luminosity limit of the host galaxies is applied to the baseline model (shown in blue). For comparison, we also show the distribution of impact parameters if no luminosity limit is applied (red, dashed line) and if a metallicity threshold is assumed (purple, dash-dotted line). Our model provides acceptable yet marginal agreement with both observed samples.
	
	Lastly, we calculate the fraction of non-detected host galaxies given the adopted luminosity limit and redshift range ($2 < z < 3.5$). From our model, we find that 42~\% of GRB host galaxies fall below the adopted luminosity limit for detection of the host galaxies. There is, however, a significant uncertainty ascribed to this estimate given the many assumptions mentioned above. We can gauge this uncertainty by varying the luminosity limit. A change of $\pm$1~mag corresponds to a range of 35--49~\% of undetected hosts.


\section{Discussion}
\label{discussion}
	
	The fair agreement between our baseline model and the observed impact parameters and metallicities of $2 < z < 3.5$ GRB-DLAs suggests that GRBs sample the full luminosity function of star-forming galaxies weighted by their star-formation rate. This agreement would indicate that GRB-DLAs and quasar-DLAs sample the luminosity function of star-forming galaxies in the same way. This is discussed further in Sect.~\ref{discussion:dlas}. Nonetheless, there are significant differences in the absorption properties between quasar- and GRB-DLAs. Most notably the distributions of \NHI\ and \Av, see middle and lower panels of Fig.~\ref{fig:CDF}.
	The higher average \NHI\ is not straight-forward to explain by our baseline model, and requires the inclusion of one additional free parameter, namely the over-density of neutral gas, either from gas around the GRB progenitor or in the GRB host ISM in general. We will explore this scenario in more detail in Sect.~\ref{discussion:overdensity}.
	However, we emphasize that the increased gas density would not directly affect the distribution of impact parameters unless over-dense star forming regions only occur at specific galactic radii which we deem rather unlikely. Similarly for the metallicity, the increased density would not affect the metallicity of the gas. The added \NHI, however, affects the distributions indirectly through the higher fraction of GRBs that are missed due to optical extinction by dust. Our conclusion that GRB hosts should represent the underlying population of star-forming galaxies at $z>2$ therefore still holds regardless of the additional density parameter introduced in this work. Nonetheless, not all GRB hosts are detectable at current luminosity limits, see the discussion in Sect.~\ref{discussion:dark}.

	While our model is consistent with the data without including a metallicity threshold, we cannot firmly exclude the possibility of a metallicity threshold mainly due to the low number statistics but also due to the redshift range in our study. In particular, the higher metallicity threshold of $Z < 0.7Z_{\odot}$ by \citet{Palmerio2019} leaves no significant impact on our model due to the low average metallicity of galaxies at these high redshifts. The lower and more restrictive threshold put forward by \citet{Metha2020} is not ruled out either but is disfavoured by the detections of higher metallicity GRBs as well as the few large impact parameters (see Fig.~\ref{fig:impact-pars}). We caution that the metallicities presented in our analysis only probe average line-of-sight measurements and are therefore not sensitive to small-scale variations that \citet{Metha2020} discuss, or like those observed in the Milky Way \citep{DeCia2021}. Nonetheless, as the metallicity of galaxies globally increases with time \citep[see compilation by][]{DeCia2018}, such metallicity thresholds will become more apparent towards lower redshifts than what we study in this work \citep[$z<2$; see also][]{Perley2013, Perley2016b, Palmerio2019}, for which we do not have absorption data in this work due to the need for space-based spectra to constrain \NHI\ and thus the metallicity.

	\subsection{Over-density of neutral hydrogen in GRBs}
	\label{discussion:overdensity}
	
	Our modelling indicates that the average CNM density is 9 to 14 times higher for GRB sight lines compared to quasar-DLAs with CNM tracers (\CI\ or H$_2$; see Sect.~\ref{sect:model-corrections}), which in our model arise at similar impact parameters.  GRB sight lines (and thus GRB-DLAs) are mainly associated with the death of massive stars. We therefore propose two plausible explanations for the observed over-density: 1) the excess \HI\ is cased by material locally around the massive progenitor, or 2) the host-galaxy is overall exhibiting an increase in ISM density, which could be expected during a burst of star formation. In the following, we will explore both of these scenarios in more detail.

	\subsubsection{Are GRBs tracing specific over-dense regions of the ISM?}
	
	Massive stars are only formed in very dense gas clouds. GRBs could therefore be biased tracers of massive, over-dense regions of the ISM. Such over-densities around GRBs are also inferred from X-ray absorption \citep[e.g.,][]{Watson2007, Schady2011}, in which case the over-density is mainly ascribed to an excess of ionized material along the line of sight. However, the X-ray absorption column density is found to correlate with the neutral gas absorption \citep{Watson2013} which would further support the idea that GRBs could probe over-dense regions of the ISM.
	
	The GRB explosion itself may also affect the neutral gas around the progenitor star. However, photo-ionization modelling indicates that ionization from the burst itself has limited effects on the \HI\ column density beyond a few pc from the burst\footnote{The effects may be significant out to $\sim$100~pc if the initial column density is lower but this would argue against the {\it higher} column densities observed on average. Moreover, the models presented by \citet{Ledoux2009} do not consider the shielding from dust which would further reduce the ionizing effects of the burst.} \citep{Ledoux2009}. We therefore do not consider this a significant effect to the full line-of-sight column density given the inferred distances between burst and bulk absorption of the order $\sim$100\,pc but may be up to 1.7\,kpc \citep{Vreeswijk2007, Vreeswijk2013, DElia2011, Hartoog2013}. The effect of photo-ionization of the molecular gas phase will be discussed in Sect.~\ref{discussion:cnm}.
	
	The most massive stars, thought to give rise to GRBs, will explode very early in the star formation process. If the progenitor is more massive than $\sim30~{\rm M}_{\odot}$, the lifetime will be $\lesssim$5--10~Myr \citep*{Woosley2002}. Simulations show that the dense molecular clouds rarely survive more than a few Myr due to intense radiative feedback from the massive stars \citep{Kimm2022}. However, in a few cases, depending on geometry and density, the clouds may survive up to 5~Myr. Recent work looking at high-resolution simulations of a dwarf galaxy shows that such massive birth clouds may re-accrete after the initial dispersion by radiative feedback, extending the lifetimes of these clouds up to the order of 10~Myr \citep{Jeffreson2024}.

\begin{figure}
	\includegraphics[width=0.5\textwidth]{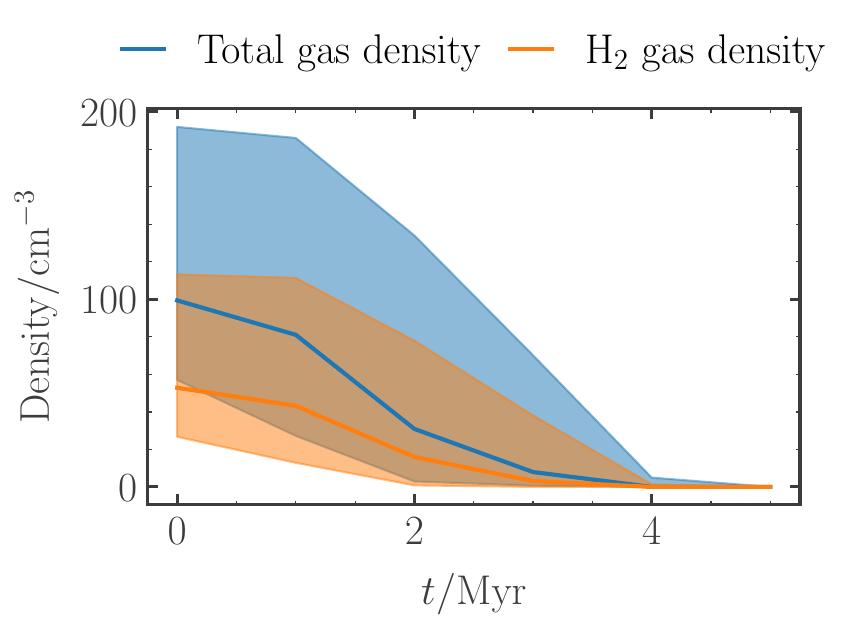}
	\caption{Median total gas volume density (blue) and molecular hydrogen volume density (orange) as a function of time after stellar birth $t$, for each star particle of mass $\gtrsim 30$~Myr (H/He-burning lifetime $\leq 6~Myr$, see text) in a simulation of a dwarf spiral galaxy.}
	\label{fig:density-evolution}
\end{figure}

	We estimate the time-dependent median gas volume density and H$_2$ gas volume density around massive, GRB-progenitor stars formed within dense molecular clouds, using the output from a high-resolution, chemo-dynamical simulation of an entire dwarf spiral galaxy \citep[see][]{Jeffreson2024}. The mass resolution of the simulation is $859~{\rm M}_\odot$, and so all stellar populations are synthesized for star clusters (stellar particles) of median mass 859~${\rm M}_\odot$, drawn from a \citet{Chabrier2003} initial mass function.

	Because the run-time memory required to store the stellar population for each star particle in the high-resolution simulation is prohibitively-large, we do not have access to the masses of individual stars that were formed during the simulation by \citet{Jeffreson2024}. Instead, a supernova progenitor mass of $>30$~M$_{\odot}$ is assumed corresponding to a lifetime of $<6$~Myr \citep[e.g.,][]{Chabrier2003}. We therefore select all stellar particles in the simulation that produce supernovae within 6 Myr of their formation, and count the number of supernovae $N_{\rm SN}$ that occur in each particle before this time. We then compute the total gas density $\rho$ and molecular fraction $x_{\rm H_2}$ within a sphere of radius 18~pc (similar to a giant molecular cloud) around each selected stellar particle, during each Myr before its death. In the simulation, there are $5 \times 10^4$ such massive progenitor stars over a period of 300 Myr. The $N_{\rm SN}$-weighted median values of $\rho$ and $\rho_{\rm H_2}$ across all selected star particles over this 6 Myr interval, along with their interquartile ranges, are shown in Fig.~\ref{fig:density-evolution}. 

	If we assume a standard average gas density of $\sim$1--10~cm$^{-3}$ \citep{Ferriere2001}, our modelling suggests a gas density around GRBs of the order 10--150~cm$^{-3}$.
	This would imply an average lifetime of the GRB progenitor of $<3$~Myr based on the results shown in Fig.~\ref{fig:density-evolution}. Such a short lifetime is still consistent with expected lifetimes of massive stars \citep{Woosley2002}; However, we find it unlikely that the re-accretion of only the progenitor birth cloud is enough to explain the over-density that we infer for GRB sight lines compared to quasar-DLAs with CNM tracers. This is further supported by the rather large inferred distances between explosion site and the bulk of the gas giving rise to fine-structure excitation of metal lines \citep{Vreeswijk2007, Vreeswijk2013, DElia2011}. Hence, the bulk of the absorption most likely does not arise from the immediate surroundings of the GRB progenitor.

	\subsubsection{Are GRBs tracing the onset of a burst in star formation?}

	Another interpretation of the over-density of \HI\ would be that GRBs probe galaxies at specific {\it times} in their star-formation histories. That is, if GRBs mainly occur at the onset of a star-burst, or shortly thereafter, the host galaxy may contain more such over-dense star-forming clouds on average. This is consistent with observations of local star-bursts whose overall gas surface densities are 1--2 orders of magnitude higher than regular galaxies \citep[e.g.,][]{Kennicutt1998, Bigiel2008}. Such an over-density would shift the radial column density profile used in our model (Fig.~\ref{fig:radial_logN}) towards higher \logNHI\ for GRBs. Our modelling is consistent with the observations without any further changes to the slope of the radial column density distribution. The slight excess of GRB-DLAs with $\logNHIcm \sim 21$ compared to our model (see Fig.~\ref{fig:CDF}) could be explained by a slightly flatter radial dependence. However, we caution that a full analysis of the radial distribution of \logNHI\ would require more sophisticated models. In future work, we plan to investigate the radial \logNHI\ profile in more detail assisted by numerical simulations.

	In this temporal interpretation, the GRB sight lines are thus more likely to intercept high-density neutral gas compared to other times where such dense gas has been dispersed by radiative feedback from the embedded, young stars. A similar conclusion is reached by \citet{Hatsukade2020} who study the molecular gas mass of GRB host galaxies traced by CO emission. These authors find that the GRB host galaxies have higher molecular gas mass fractions than regular star-forming galaxies; However, this excess of molecular gas disappears when comparing to galaxies that have higher specific star-formation rates similar to GRB hosts. 

	The temporal interpretation of the over-density is also consistent with the enhanced specific star formation rates often observed in GRB host galaxies \citep{Christensen2004, Perley2013, Bjornsson2019}. If GRBs predominantly trace a specific time shortly after the onset of a starburst \citep[see][]{Hatsukade2020}, the instantaneous SFR observed would be elevated compared to a random galaxy selected at a more representative phase in its star-formation history such as quasar-DLAs with CNM tracers. A logarithmic offset in column density would then naturally follow from the power-law relation between star-formation rate density and gas density \citep{Schmidt1959, Kennicutt1998}. We therefore favour this interpretation of GRBs as tracers of a specific moment in the host galaxy's star formation history.

	\subsection{The fraction of dark bursts}
	\label{discussion:dark}
	The fraction of so-called `dark bursts' is still a subject of debate, and several definitions exist based primarily on X-ray classification \citep[e.g.,][]{Jakobsson2004, vanderHorst2009} or optical dust obscuration \citep[e.g.,][]{Kruhler2011, Melandri2012}. The fraction of dark bursts is estimated to be around 10--20~\% \citep{Jakobsson2004, vanderHorst2009, Perley2013} but may be up to 30--40~\% \citep{Greiner2011, Melandri2012} for an adopted dark burst definition of $\Av > 1$~mag. Using this definition, our GRB model predicts a dark burst fraction of 29\%, which is broadly consistent with the GRB observations. Yet, our model predicts a larger fraction of highly obscured bursts ($\Av > 3$~mag) compared to the sample by \citet{Covino2013}. We attribute this larger fraction of dark bursts in our model to the fact that some GRBs are expected to be so obscured that no optical or near-infrared afterglow is identified \citep{Fynbo2009}. These highly obscured bursts would therefore not be present in the sample by \citet{Covino2013} as these authors only include GRBs with a spectroscopic redshift determination. However, these highly obscured GRBs with no detected afterglow may also be caused by other factors, such as being at very high redshift or intrinsically faint bursts. It is therefore not straightforward to quantify the fraction of GRBs with no optical/near-infrared afterglow due to dust obscuration alone.

	A related issue to the one of dust-obscured dark GRBs is the number of host galaxies that are not detected \citep[see][for a complete sample of GRBs]{Hjorth2012}. These undetected host galaxies may bias the associations of impact parameters \citep{Blanchard2016, Lyman2017}.
	In our model we have tried to account for this undetected population of galaxies given the estimated limiting luminosity. We find that our estimated fraction of undetected host galaxies of 35--49~\% agrees well with the 40~\% of non-detected hosts by \citet{Lyman2017}. However, the observations are based on inhomogeneous imaging data and varying detection limits in the two samples that we compare to.
	\citet{Hjorth2012} study a more complete sample and find a smaller fraction of about 20\%.
	A fair comparison to our GRB model is further complicated due to the differences in the modelled rest-frame UV luminosities as compared to the observed rest-frame optical range. Significant variations are observed between the line-of-sight dust extinction and the global attenuation of the host galaxy \citep{Kruhler2011, Friis2015, Heintz2017, Chrimes2019, Schroeder2022}. These variations are most likely due to an inhomogeneous dust distribution and possibly very localized dust near the burst \citep{Greiner2011, Kruhler2011}. However, on average there seems to be a tendency for more dusty host galaxies among the most dust-obscured GRB afterglows \citep{Perley2013, Corre2018, Schroeder2022}. These complications in dust-corrections may explain the slight tension we observe in the predicted impact parameter distributions. From an observational point of view, it is also not straightforward to associate a host galaxy to a GRB as evidenced by the significant differences between the two samples of \citet{Blanchard2016} and \citet{Lyman2017}.

	\subsection{CNM tracers in GRB sight lines}
	\label{discussion:cnm}
	
	Regardless of whether GRBs arise in specific over-dense pockets in the CNM or at specific times of star-burst activity, one would naively expect to observe tracers of the CNM such as \CI\ or H$_2$ in the majority of GRB sight lines. However, only around 30--40\% of GRBs have CNM tracers detected by either H$_2$ absorption \citep{Bolmer2019} or \CI\ absorption \citep{Heintz2019a}. Indeed, our statistical model predicts a detection rate of nearly 100\% based on the detection limit of $\log(N_{{\rm H}_2}/{\rm cm}^{-2}) \gtrsim 17$ given by \citet{Bolmer2019}.
	
	One possible explanation for the modest detection rate of CNM tracers in GRB afterglow spectra is the efficient photo-dissociation of H$_2$ and ionization of \CI\ by the GRB explosion itself \citep{Draine2002}. The ionizing flux produced by the GRB explosion may destroy H$_2$ out to $\sim$100~pc depending on the exact column density \citep{Ledoux2009}.
	\citet{Whalen2008} on the other hand find that the GRB explosion itself is not able to fully destroy all H$_2$ near the GRB. They argue that H$_2$ must already be suppressed before the GRB goes off. The simulations by \citet{Jeffreson2024} show that H$_2$ can indeed be significantly suppressed by radiative feedback around massive stars, thereby lowering the column density of H$_2$. If the column density is already suppressed before the burst, the GRB explosion will more effectively destroy the molecular gas phase \citep{Ledoux2009}. If we include this destruction of H$_2$ within 100~pc in our model, we find an H$_2$ detection fraction of 90\%. This is still inconsistent with the observations. The distance out to which H$_2$ is destroyed would need to be greater than 500~pc in order to match the moderate detection rate reported by \citet{Bolmer2019} and \citet{Heintz2019a}.
	
	Another explanation is related to the assumed azimuthally symmetric geometry. While this works well for the extended \HI\ gas, the CNM bearing gas might realistically be confined to a more flattened, disk-like structure. GRB sight lines arising from these CNM regions are then more likely to be perpendicular to the disk rather than piercing through it. This would effectively lower the volume of the CNM-bearing medium thereby lowering the probability of intercepting a CNM cloud. However, this would go against the excess of \NHI\ if the column density is not dominated by the environment close to the GRB, as discussed in Sect.~\ref{discussion:overdensity}. In this flattened CNM geometry, the probability of crossing other high-density star-forming regions would be lower, leading to the same \NHI\ distribution as for CNM-bearing DLAs, which we do not observe. Hence, unless the local environment of the GRB contributes significantly to the observed excess of \NHI, or the \HI\ distribution is different all together, we find this purely geometrical interpretation hard to reconcile with the data. Since our model is not based on a full 3-dimensional geometry, a full test of the geometry is beyond the scope of this work. In future work, we will address such higher order geometrical considerations assisted by state-of-the-art numerical simulations.

	Lastly, our model assumes a uniform covering fraction of the CNM gas for random background sources piercing the {\it full volume} of CNM-bearing gas. Since the volume-filling factor of CNM clouds may be quite low \citep[e.g.,][]{Krogager2018b}, we do not expect sight lines arising from within the CNM-bearing volume to hit a cold gas cloud in all cases. Instead, the probability of piercing another cold gas cloud is proportional to the fraction of the volume through which the sightline travels (for uniformly distributed clouds). The GRB explosion may therefore locally destroy H$_2$ (or \CI) but the absorption sightline still propagates through the rest of the host galaxy ISM. If the GRBs are randomly distributed within the volume where cold clouds are found, the average detection fraction of CNM tracers in our model would instead be $\sim$50~\%. 
	
	The projected covering fraction of CNM gas in quasar absorption systems may in fact be closer to $\sim$80\% \citep{Wiklind1995, Wiklind2018, Krogager2018b, Boisse2019}. An overall lower projected covering fraction would further decrease the chance of a random GRB sightline piercing a cold cloud, bringing the expected H$_2$ detection rate into agreement with the observations \citep[30-40\%][]{Bolmer2019, Heintz2019a}. We therefore conclude that the most likely explanation for the low detection rate of H$_2$ and \CI\ is due to a combination of local suppression of these species due to the prompt emission of the GRB and a low volume-filling factor of cold gas clouds in the host galaxy ISM.

	\subsection{How do quasar-DLAs and GRB-DLAs trace star-forming galaxies?}
	\label{discussion:dlas}
	
	There is a fundamental relationship between the velocity width of quasar-DLA gas and the metallicity of the same gas \citep{Ledoux2006}; This relationship has a well-defined and clear evolution with cosmic time which has been traced back to $z=5.1$ \citep{Moller2013}. If GRBs, at all redshifts, randomly select galaxies from the same population of galaxies as do quasar-DLAs, then the resulting sample should follow the same relation with the same evolution.
	\citet{Arabsalmani2015} have tested and confirmed that the two samples for GRB- and quasar-DLAs are indeed consistent with being drawn from the same relation (see the lower panel of their fig. 2), providing strong support for the validity of our combined model assumption in this work. \citet{Arabsalmani2015} also find that while following the same relation, GRB-DLAs preferentially populate the high metallicity end of the relation suggesting that GRB selection is weighted either towards higher metallicity galaxies, towards higher metallicity sight lines in the same galaxies as quasar-DLAs, or both. \citet{Christensen2014} find that DLA hosts have negative metallicity gradients from their centre and out, meaning that a GRB sightline (located close to the centre) in a DLA galaxy would indeed be expected to show higher metallicity than a random quasar-DLA sightline.

\begin{figure}
		\includegraphics[width=0.48\textwidth]{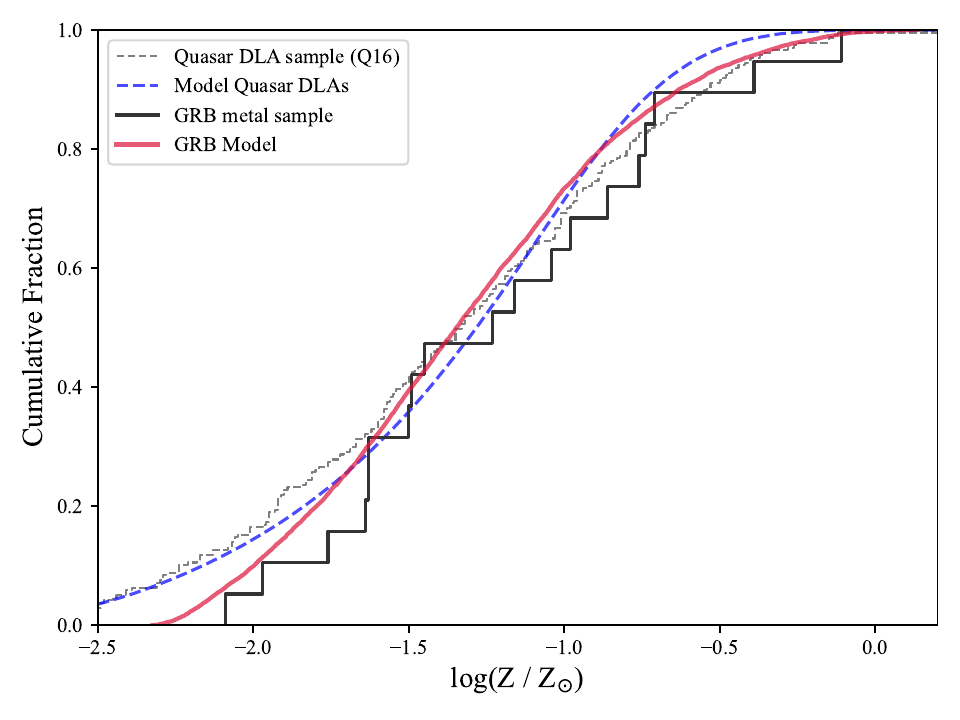}
		\includegraphics[width=0.48\textwidth]{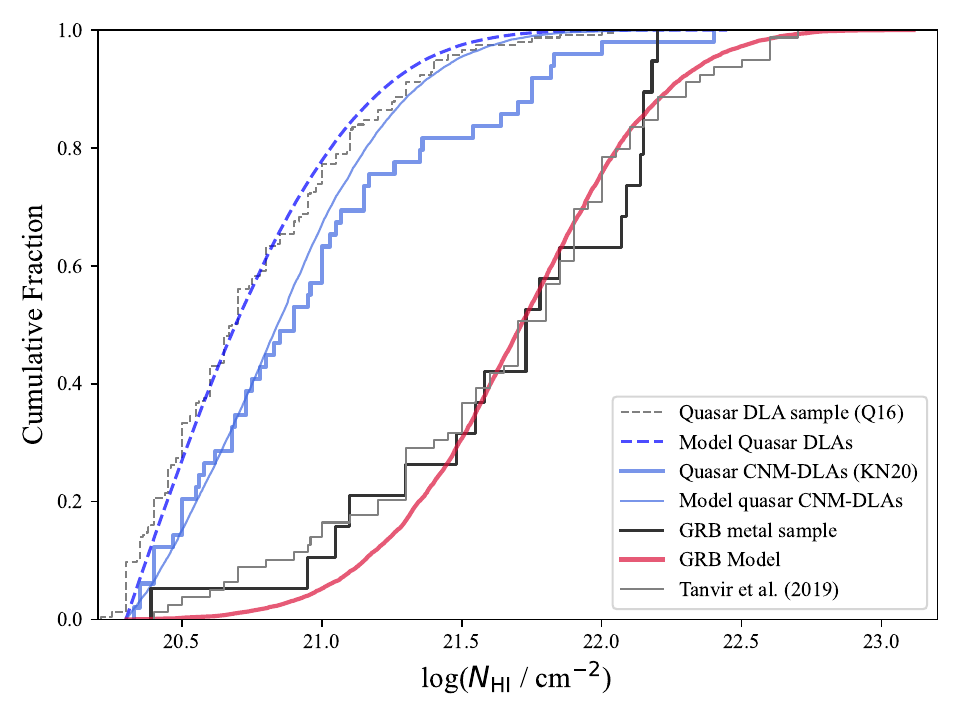}
		\caption{Cumulative distributions of metallicity (top) and \NHI\ (bottom) for the GRB metal sample shown by the solid black line and for quasar DLAs \citep[Q16]{Quiret2016} shown by the dashed gray line. We show the model distributions for GRBs in the solid, red line, and for quasar DLAs in the dashed, blue line. For comparison, we show the model distributions of quasar CNM-DLAs (i.e., with \CI\ or H$_2$ absorption) as the thin, light blue line and the observed distribution of \NHI\ for quasar CNM-DLAs \citep[KN20]{Krogager2020b} as the thick, light blue step-wise line. We do not show the metallicity distribution for the KN20 sample as this sample is biased towards high metallicity by being mostly selected based on \CI\ which favours high metallicity \citep{Ledoux2015}.
		}
		\label{fig:dla_comparison}
	\end{figure}

\begin{figure}
		\includegraphics[width=0.48\textwidth]{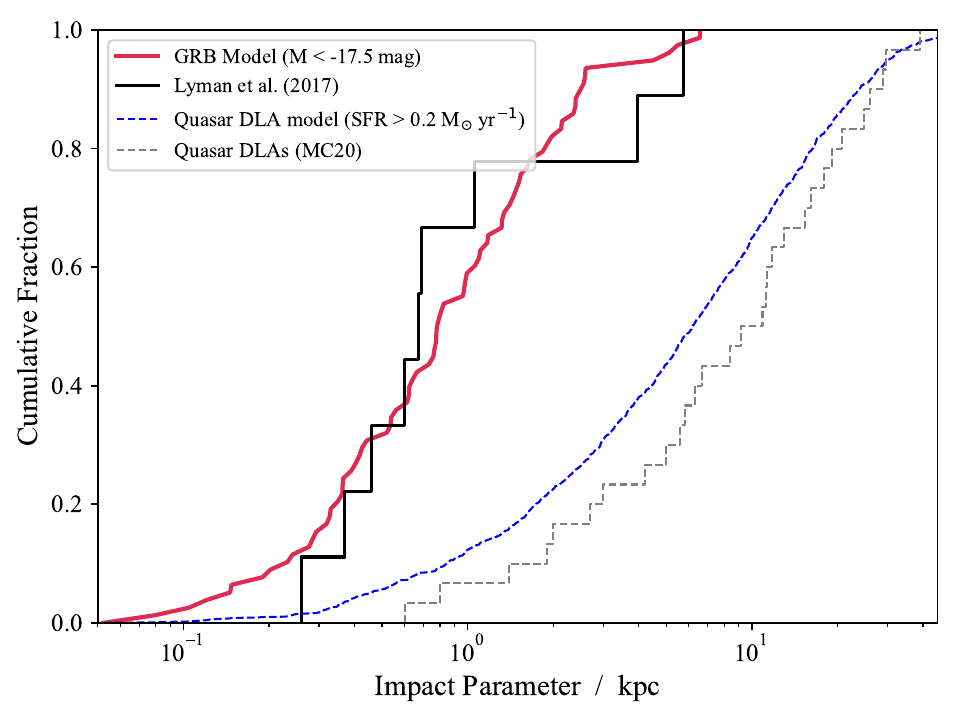}
		\caption{Cumulative distributions of impact parameters for the GRB sample by \citet{Lyman2017} shown by the solid black line and for quasar DLAs \citep[MC20]{Moller2020} shown by the dashed gray line. We show the model distributions for GRBs in the solid red line (restricted in luminosity and non-obscured bursts, see Sect.~\ref{results}). For quasar DLAs we show the model as the dashed blue line (restricted in star-formation rate based on UV luminosity to mimic observational limitations at these high redshifts, see also \citep{Krogager2020}).
		}
		\label{fig:dla_comparison_b}
	\end{figure}

	In agreement with the works mentioned above, our modelling supports the hypothesis that DLAs observed in both quasar and GRB sight lines trace the same underlying population of galaxies. On the one hand, quasar-DLAs trace galaxies weighted by their cross-section of neutral gas, which we assume is directly proportional to luminosity \citep{Krogager2020}. On the other hand, GRB-DLAs trace galaxies weighted by the star-formation rate which is also assumed to scale with luminosity \citep{Kennicutt1998}.
	This is slightly at odds with the previous work \citep{Fynbo1999,Fynbo2008} who find that the quasar-DLA cross-section scales with a power of $0.8$ instead of a power of $1$ that we adopt in this work. This change in the scaling of DLA cross-section as a function of luminosity is related to the updated metallicity measurements used by \citet{Krogager2020} as well as the direct modelling of a dust-obscuration bias which alters the metallicity distribution.
	To illustrate this, we show a comparison of the metallicity distributions for quasar-DLAs and GRB-DLAs in Fig.~\ref{fig:dla_comparison}. The quasar-DLAs extend to lower metallicities given the metallicity gradient and their larger impact parameters on average (see Fig.~\ref{fig:dla_comparison_b}). We find that the median of the GRB-DLA metallicities is 0.2~dex higher than for the quasar-DLAs after applying the correction for the redshift evolution in the mass--metallicity relation \citep{Moller2013}, in agreement with the offset reported by \citet{Arabsalmani2015}.
	
	In Fig.~\ref{fig:dla_comparison}, we also compare the \NHI\ distributions of quasar-DLAs and GRB-DLAs. We interpret the significant offset as a result of GRBs tracing their host galaxies at the onset of a star formation event leading to an overall increase in gas surface density in the host galaxy. Quasar-DLAs, by selection, trace fully random sight lines through the DLA cross-section of the galaxies at no specific location nor time. We note that the \NHI\ distribution of quasar CNM-DLAs compiled by \citep{Krogager2020b} shows a marginal excess of high \NHI\ systems. We caution, however, that the sample is not homogeneously selected and is therefore not fully representative. Indeed, the sample includes a number of targets that have been selected based on having $\logNHIcm \gtrsim 21.5$ \citep{Noterdaeme2014, Ranjan2020}. Moreover, given the low number of CNM-DLAs, a KS-test reveals that this excess is not significant given the $p$-value of 0.22.
	
	Lastly, we compare the impact parameter distributions between quasar-DLAs and GRB-DLAs in Fig.~\ref{fig:dla_comparison_b}. Our model has been restricted to only consider galaxies that would be bright enough for a detection in ground-based observational campaigns (see Sect.~\ref{results} for GRB-DLAs). For quasar-DLAs we estimate the SFR from the model UV luminosity and consider only hosts with SFR~$>0.2$~M$_{\odot}~{\rm yr}^{-1}$, see also \citet{Krogager2020}.  It is clearly seen that quasar-DLA sight lines on average probe the outer regions of their host galaxies. The impact parameters of quasar-DLAs are one order of magnitude larger than those of GRB-DLAs. We further observe that the impact parameters of quasar-DLAs are under-predicted by $\sim$20~\% in our model. This slight offset, however, is due to the complications of detecting the galaxies at small projected separations from the bright background quasar (less than $\sim 5$~kpc). Though we caution again that the impact parameter sample may further be biased by inhomogeneous observations often targeting metal-rich absorbers \citep{Fynbo2010, Krogager2017} which in our model would also be larger and are thus more likely to have larger impact parameters.

\section{Summary}
\label{summary}

	In this work, we study how GRB-DLAs probe their host galaxies using a statistical modelling approach. We compare observed distributions of \NHI, metallicity, \Av, and impact parameters to the statistical model by \citet{Krogager2020b}. We take into account a geometric correction since GRBs arise from within the volume of gas instead of piercing the full volume as a quasar sightline would. We find that this `baseline model' provides an acceptable fit to the metallicity and impact parameter distributions of GRB-DLAs. These are the more fundamental quantities related to the host-galaxy luminosity (or mass). Yet, the `baseline model' does not reproduce the observed \NHI\ nor \Av\ distributions.
	We therefore include one additional free parameter to model the over-density of neutral gas observed in GRB-DLAs \citep{Heintz2019a}. Lastly, we change the implemented dust bias prescription, which was tailored to model quasar-DLAs.
	
	We fit the dust-bias parameter and the \HI\ over-density parameter simultaneously to a sample of GRBs detected in the redshift range $2 < z < 3.5$ (Sect.~\ref{data}). From this fit, we find that bursts with $\Av > 0.44\pm0.05$~mag are too heavily obscured to allow optical spectroscopy needed to determine the metallicity\footnote{The parametrization of the dust-obscuration is of course overly simplified, as metallicity measurements could still be possible at higher levels of obscuration depending on the brightness of the GRB. Nonetheless, the parameter is useful for the statistical comparison.}. The over-density of neutral gas along the GRB sightline is found to be $\Delta \logNHI = 1.06 \pm 0.09$ (a factor of 9 to 14), see also Fig.~\ref{fig:dla_comparison}.

	The over-density of \HI\ is interpreted as a temporal selection effect, namely that GRBs, due to their massive stellar progenitors, select galaxies in an early stage of a star-formation event or a star-burst. We speculate that galaxies in such a stage of a star-burst may host many such massive and dense regions which would increase the column of gas along the line of sight \citep[][]{Hatsukade2020}. Furthermore, local starburst galaxies exhibit an excess surface density of neutral gas compared to more regular galaxies \citep[e.g.,][]{Kennicutt1998, Bigiel2008}. A similar effect could thus plausibly be expected for high-redshift galaxies as well. This interpretation is also qualitatively in agreement with the claim that GRB host galaxies have higher specific star-formation rates on average \citep{Christensen2004, Perley2013, Bjornsson2019}. We find it less convincing that the over-density of neutral gas is caused by gas near the progenitor itself, unless the lifetime of the progenitor is $<3$~Myr and the majority of the progenitors are able to re-accrete significant amounts of their birthclouds as suggested by \citet{Jeffreson2024}.
	
	Based on the acceptable agreement between metallicity and impact parameters from the model, both before and after fitting the \NHI\ and \Av\ distributions, we conclude that GRB-DLAs and quasar-DLAs with CNM tracers sample the luminosity function of star-forming galaxies in the same way. We further test the metallicity thresholds put forward in the literature and find that at these high redshifts ($z>2$), such metallicity thresholds do not significantly affect the GRB population due to the overall lower metallicities at early times in the Universe. The host galaxies of GRB-DLAs are therefore representative of the underlying population in terms of metallicity and luminosity (or stellar mass), yet still susceptible to biases due to optical dust obscuration.
	
	We quantify the effect of dust obscuration and compare the fraction of so-called `dark' bursts (with $\Av > 1$~mag) to our model predictions. Observations infer a rather uncertain range of dark bursts ranging from 10 to 40\% \citep{Jakobsson2004, Melandri2012}. Based on our model, we expect 29\% of bursts to have $\Av>1$~mag.
	
	Under the assumption that GRBs arise from the inner regions of galaxies where cold neutral gas should be present, we would naively expect that all GRB sight lines should show tracers of the CNM (\CI\ or H$_2$). We discuss the detection rate of such CNM tracers and find that the observed fraction of $30-40$~\% \citep{Bolmer2019, Heintz2019a} is consistent with our model, see Sect.~\ref{discussion:cnm}.
	
	Lastly, we compare the observed metallicity distribution of quasar- and GRB-DLAs to the expected distributions from our model framework. Quasar-DLAs have slightly lower metallicities on average due to the metallicity gradient and much larger impact parameters (Fig.~\ref{fig:dla_comparison_b}). The GRB-DLAs have metallicities that are 0.2~dex higher on average, consistent with \citet{Arabsalmani2015}, when taking into account the redshift evolution in the mass--metallicity relation \citep{Moller2013}.
	In contrast, the quasar-DLAs have significantly lower \NHI\ on average, and even when comparing to the subset of quasar-DLAs with CNM gas that probe similar small impact parameters as GRB-DLAs, the median \NHI\ differs by almost an order of magnitude. This bolsters our conclusion that the observed excess of \NHI\ in GRB-DLAs is not simply due to the higher gas density at smaller galactic radii (see Figs.~\ref{fig:dla_comparison} and \ref{fig:dla_comparison_b}).

\section*{acknowledgements}
	The authors would like to thank the referee for the very constructive and thorough review of this work.
	JKK would like to thank the ESO Visitor Programme for the financial support to visit ESO Garching in April 2023 which allowed us to focus on this project and make significant progress on the article.
	JKK and ADC acknowledge support from the Swiss National Science Foundation under grant 185692. KEH acknowledges support from the Carlsberg Foundation Reintegration Fellowship Grant CF21-0103. LC is supported by the Independent Research Fund Denmark (DFF 2032-00071). SMRJ is supported by Harvard University via the ITC Fellowship. JPUF is supported by the Independent Research Fund Denmark (DFF--4090-00079) and thanks the Carlsberg Foundation for support.
	The Cosmic Dawn Center (DAWN) is funded by the Danish National Research Foundation under grant No. 140.

\section*{data availability}
	All data analysed in this article are collected from literature and are referenced throughout the text. The main data underlying this analysis are available in Table~\ref{tab:metals}. Additional data are available in the refereed works cited in this article.

\bibliographystyle{mnras}

\end{document}